\documentclass{PoS}

\title{Diffractive exclusive production of heavy quark pairs
at high energy proton-proton collisions}

\ShortTitle{Exclusive production of heavy quark pairs}

\author{\speaker{Antoni Szczurek}\\
        Institute of Nuclear Physics PAN, PL-31-342 Cracow,
        Poland,\\
        Univerisity of Rzesz\'ow, PL-35-959 Rzesz\'ow, Poland \\
        E-mail: \email{Antoni.Szczurek@ifj.edu.pl}}

\author{R. Maciu{\l}a $^1$, R. Pasechnik $^{2}$ \\
        $^1$ Institute of Nuclear Physics PAN, PL-31-342
        Cracow,Poland,\\
       $^2$ High Energy Physics, Department of Physics and Astronomy, 
       Uppsala University Box535, SE-75121 Uppsala, Sweden}

\abstract{
We discuss exclusive double diffractive (EDD) production 
of heavy quark - heavy antiquark pairs
at high energies. Differential distributions for $c \bar c$ 
at $\sqrt{s}$ = 1.96 GeV and for $b \bar b$ at $\sqrt{s}$ = 14 TeV are
shown and discussed. 
Irreducible leading-order $b \bar b $ background to Higgs production is calculated
for the first time in several kinematical variables.
The signal-to-background ratio is shown and several improvements are suggested.
}

\FullConference{XVIII International Workshop on Deep-Inelastic Scattering and Related Subjects, DIS 2010\\
		April 19-23, 2010\\
		Firenze, Italy}

\begin{document}

\section{Introduction}

There is recently a growing theoretical interest in studying exclusive
processes. Exclusive production of the Higgs boson is a flag process of special
interest and importance. Only a few processes have been measured so far
at the Tevatron (see \cite{Albrow} and references therein).
Khoze, Martin and Ryskin developed an approach in the language of
off-diagonal unintegrated gluon distributions.
This approach was applied to exclusive production of Higgs boson
\cite{KMR_Higgs}. 
In our recent paper we applied the formalism
to exclusive production of $c \bar c$ quarks. Quite large cross sections
have been found \cite{MPS10}. 

The cross section for the Standard Model Higgs production is of the order of
1 fb for $M_H$ = 120 GeV \cite{KMR_Higgs}. The dominant $b \bar b$ decay
channel is therefore preferential from the point of view of statistics.
The $b \bar b$ exclusive production was estimated only at higher order
\cite{KMR_bbbar_suppression}.
It was argued that the leading-order contribution is rather small
using a so-called $J_z$ = 0 rule \cite{KMR_bbbar_suppression}. Here we
show a quantitative calculation which goes beyond this simple
rule. In our calculation we include exact matrix element for massive
quarks and the 2 $\to$ 4 phase space.
This fully four-body calculation allows to impose cuts on any kinematical
variable one wish to select. Different types of backgrounds to Higgs
production were studied before e.g. in Ref.\cite{CHHKP09}.

\section{Formalism}

\subsection{The amplitude for $p p \to p p Q \bar Q$}

Let us concentrate on the simplest case of the production of $q\bar{q}$
pair in the color singlet state. Color octet state would demand an
emission of an extra gluon \cite{KMR_bbbar_suppression} which considerably
complicates the calculations. We do not consider the $q \bar q g$
contribution as it is higher order compared to the one considered here.

In analogy to the Khoze-Martin-Ryskin approach (KMR)
\cite{KMR_Higgs} for Higgs boson production, we write
the amplitude of the exclusive diffractive $q\bar{q}$ pair
production $pp\to p(q\bar{q})p$ in the color singlet state as
\begin{eqnarray}
{\cal
M}_{\lambda_q\lambda_{\bar{q}}}^{p p \to p p q \bar q}(p'_1,p'_2,k_1,k_2) &=&
s\cdot\pi^2\frac12\frac{\delta_{c_1c_2}}{N_c^2-1}\,
\Im\int d^2
q_{0,t} \; V_{\lambda_q\lambda_{\bar{q}}}^{c_1c_2}(q_1, q_2, k_1, k_2) \nonumber \\[5pt]
&&\frac{f^{\mathrm{off}}_{g,1}(x_1,x_1',q_{0,t}^2,
q_{1,t}^2,t_1) \; f^{\mathrm{off}}_{g,2}(x_2,x_2',q_{0,t}^2,q_{2,t}^2,t_2)}
{q_{0,t}^2\,q_{1,t}^2\, q_{2,t}^2} \; ,
\label{amplitude}
\end{eqnarray}
where $\lambda_q,\,\lambda_{\bar{q}}$ are helicities of heavy $q$
and $\bar{q}$, respectively. Above $f_1^{\mathrm{off}}$ and
$f_2^{\mathrm{off}}$ are the off-diagonal unintegrated gluon
distributions in nucleon 1 and 2, respectively. 

The longitudinal momentum fractions of active gluons
are calculated based on kinematical variables of outgoing quark
and antiquark:
$x_1 = \frac{m_{3,t}}{\sqrt{s}} \exp(+y_3)
     +  \frac{m_{4,t}}{\sqrt{s}} \exp(+y_4)$ and
$x_2 = \frac{m_{3,t}}{\sqrt{s}} \exp(-y_3)
     +  \frac{m_{4,t}}{\sqrt{s}} \exp(-y_4)$,
where $m_{3,t}$ and $m_{4,t}$ are transverse masses of the quark and
antiquark, respectively, and $y_3$ and $y_4$ are corresponding
rapidities.

The bare amplitude above is subjected to absorption corrections.
The absorption corrections are taken here in a simple multiplicative form.

\subsection{$g g \to Q \bar Q$ vertex}

Let us consider the subprocess amplitude for the $q\bar{q}$ pair production
via off-shell gluon-gluon fusion. The vertex factor
$V_{\lambda_q\lambda_{\bar{q}}}^{c_1c_2}=
 V_{\lambda_q\lambda_{\bar{q}}}^{c_1c_2}(q_1,q_2,k_1,k_2)$
in expression (\ref{amplitude}) is the production amplitude
of a pair of massive quark $q$ and antiquark $\bar{q}$ with
helicities $\lambda_q$, $\lambda_{\bar{q}}$ and
momenta $k_1$, $k_2$, respectively.
The color singlet $q\bar{q}$ pair production amplitude can be written as
\begin{eqnarray}\label{qqamp}
&&V_{\lambda_q\lambda_{\bar{q}}}^{c_1c_2}(q_1,q_2,k_1,k_2)\equiv
n^+_{\mu}n^-_{\nu}V_{\lambda_q\lambda_{\bar{q}}}^{c_1c_2,\,\mu\nu}(q_1,q_2,k_1,k_2),
\end{eqnarray}
The tensorial part of the amplitude reads:
\begin{eqnarray}\nonumber
&&{}V_{\lambda_q\lambda_{\bar{q}}}^{\mu\nu}(q_1, q_2, k_1, k_2)
= g_s^2 \,\bar{u}_{\lambda_q}(k_1)
\biggl(\gamma^{\nu}\frac{\hat{q}_{1}-\hat{k}_{1}-m}
{(q_1-k_1)^2-m^2}\gamma^{\mu}-\gamma^{\mu}\frac{\hat{q}_{1}-
\hat{k}_{2}+m}{(q_1-k_2)^2-m^2}\gamma^{\nu}\biggr)v_{\lambda_{\bar{q}}}(k_2).\\
\label{vector_tensor}
\end{eqnarray}
The coupling constants $g_s^2 \to g_s(\mu_{r,1}^2)
g_s(\mu_{r,2}^2)$. In the present calculation we take the
renormalization scale to be $\mu_{r,1}^2=\mu_{r,2}^2=M_{q \bar
q}^2/4$ or $M_{q \bar q}^2$. The exact matrix element is
calculated numerically. 

\subsection{Off-diagonal unintegrated gluon distributions}

In the KMR approach the off-diagonal parton distributions
(i=1,2) are calculated as
\begin{eqnarray}
f_i^{\mathrm{KMR}}(x_i,Q_{i,t}^2,\mu^2,t_i) &=& R_g
\frac{d[g(x_i,k_t^2)S_{1/2}(k_{t}^2,\mu^2)]}{d \log k_t^2} |_{k_t^2
= Q_{it}^2} \;
F(t_i) \;  \nonumber \\
&\approx& R_g \frac{dg(x_i,k_t^2)}{d \log k_t^2} |_{k_t^2 =
Q_{i,t}^2} \;
S_{1/2}(Q_{i,t}^2,\mu^2) \; F(t_i)  \; ,
\label{KMR-off-diagonal-UGDFs}
\end{eqnarray}
where $S_{1/2}(q_t^2, \mu^2)$ is a Sudakov-like form factor relevant
for the case under consideration. The last approximate
equalities come from the fact that in the region under consideration
the Sudakov-like form factors are somewhat slower functions of
transverse momenta than the collinear gluon distributions. While
reasonable for an estimate of gluon distribution it may be not
sufficient for precise calculation of the cross section. It is
reasonable to take a running (factorization) scale as: $\mu_1^2 =
\mu_2^2 = M_{q \bar q}^2/4$ or $M_{q \bar q}^2$.

The factor $R_g$ here cannot be calculated from first principles
in the most general case of off-diagonal UGDFs.
It can be estimated in the case of off-diagonal collinear PDFs
when $x' \ll x$ and $x g = x^{-\lambda}(1-x)^n$.
Then
$R_g = \frac{2^{2\lambda+3}}{\sqrt{\pi}}
\frac{\Gamma(\lambda+5/2)}{\Gamma(\lambda+4)}$.
Typically $R_g \sim$ 1.3 -- 1.4 at the Tevatron energy. 
The off-diagonal form factors are parametrized here as $F(t) = \exp \left( B_{\mathrm{off}} t \right)$.
In practical calculations we take $B_{\mathrm{off}}$ = 2 GeV$^{-2}$.
In the original KMR approach the following prescription for the
effective transverse momentum is taken: 
$Q_{1,t}^2 = \min\left( q_{0,t}^2,q_{1,t}^2 \right)$ and
$Q_{2,t}^2 = \min\left( q_{0,t}^2,q_{2,t}^2 \right)$.
In evaluating $f_1$ and $f_2$ needed for calculating the amplitude
(\ref{amplitude}) we use different collinear distributions.
It was proposed \cite{KMR_Higgs} to express the $S_{1/2}$ form factors in
Eq.~(\ref{KMR-off-diagonal-UGDFs}) through the standard Sudakov form
factors as:
\begin{equation}
S_{1/2}(q_t^2,\mu^2) = \sqrt{T_g(q_t^2,\mu^2)}  \; .
\label{S_{1/2}}
\end{equation}
%

\subsection{Cross section}

The cross section is calculated as
\begin{equation}
d \sigma = \frac{1}{2s} |{\cal M}_{2 \to 4}|^2 (2 \pi)^4
\delta^4(p_a + p_b - p_1 - p_2 - p_3 - p_4) \frac{d^3 p_1}{(2 \pi)^3
2 E_1} \frac{d^3 p_2}{(2 \pi)^3 2 E_2} \frac{d^3 p_3}{(2 \pi)^3 2
E_3} \frac{d^3 p_4}{(2 \pi)^3 2 E_4} \; . \label{phase_space}
\end{equation}
The details how to conveniently reduce the number of
kinematical integration variables are given elsewhere.

\section{Results}

\subsection{$pp \to pp c \bar c$ }

Let us proceed now with the presentation of differential
distributions of charm quarks produced in the EDD mechanism. In this case
we have fixed the scale of the Sudakov form factor to be
$\mu = M_{c \bar c}/2$. Such a choice of the scale leads to a strong
damping of the cases with large rapidity gaps between $q$ and $\bar q$.

\begin{figure}[h!]
\includegraphics[width=0.45\textwidth]{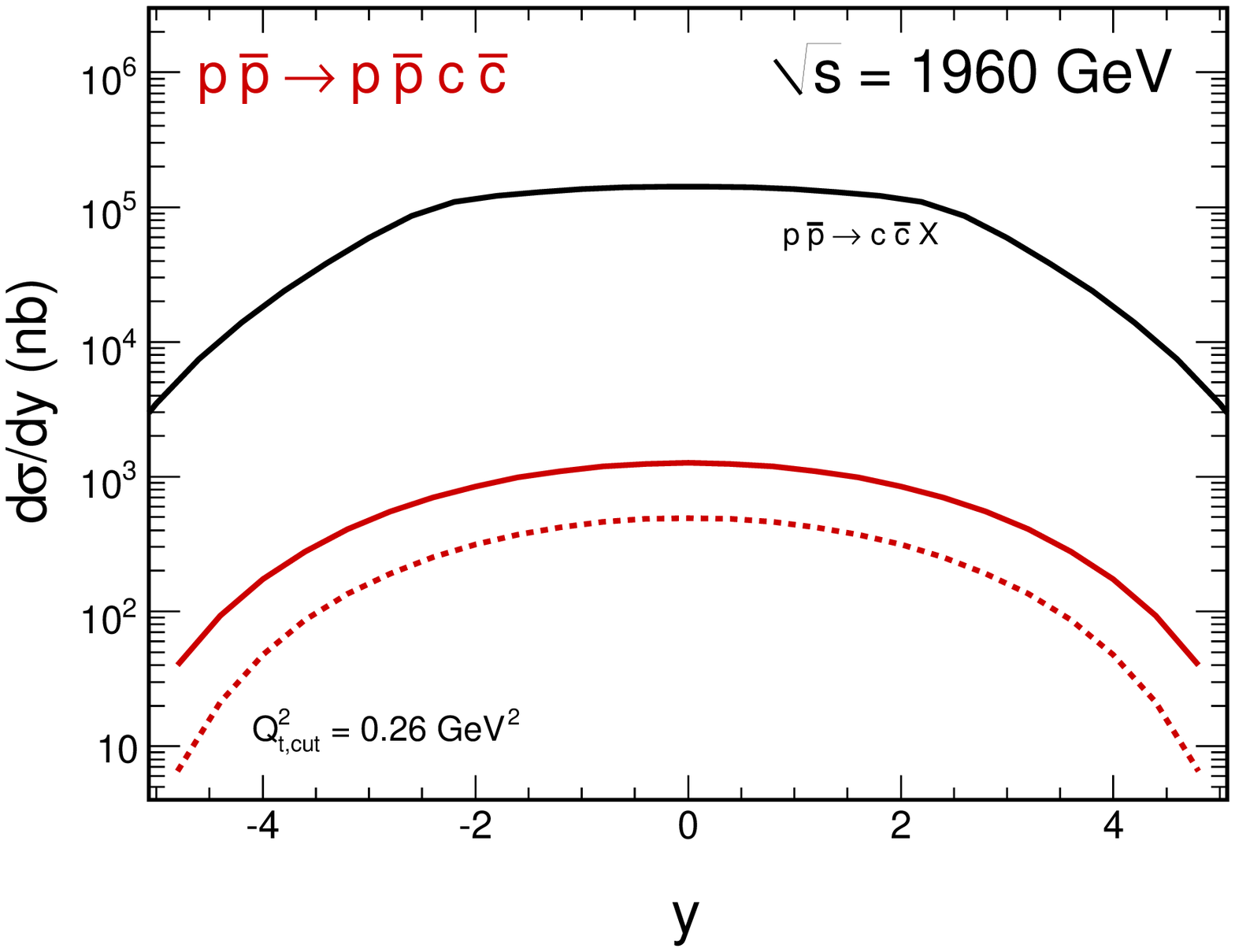}
\includegraphics[width=0.45\textwidth]{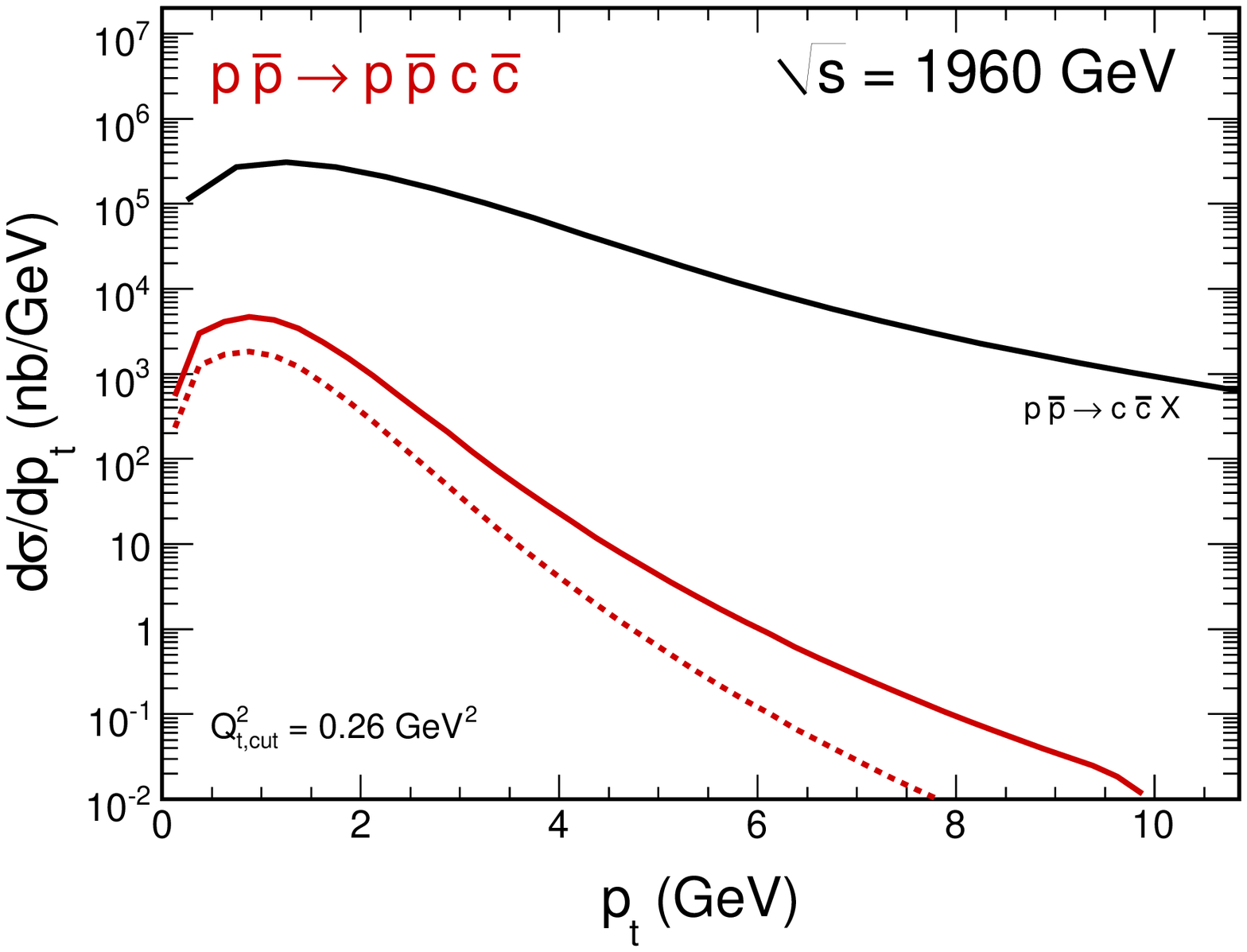}
\caption{\small Rapidity distribution of $c$ or $\bar c$ (left)
and transverse momentum distribution of $c$ or $\bar c$ (right).
The top curve is for inclusive production in the $k_t$-factorization
approach with the Kwieci\'nski UGDF and $\mu^2 = 4 m_c^2$, while the
two lower lines are for the EDD mechanism for the KMR UGDF with
leading-order collinear gluon distribution \cite{GRV94}. The solid
line is calculated from the exact formula 
and the dashed line for the
simplified formula (when only derivative of collinear GDF is taken).
An extra cut on the momenta in the loop $Q_{t,cut}^2$ = 0.26 GeV$^2$
was imposed. Absorption effects were included approximately by
multiplying the cross section by the gap survival factor $S_G$ = 0.1.
}
\label{figx}
\end{figure}

In the left panel of Fig.~\ref{figx} we show distribution in
rapidity. 
The results obtained with the KMR method are shown together with
inclusive gluon-gluon contribution.
The effect of absorption leads to a damping
of the cross section by an energy-dependent factor. For
the Tevatron this factor is about 0.1. If the extra factor
is taken into account the EDD contribution is of the order
of 1\% of the dominant gluon-gluon fusion contribution.

The corresponding rapidity-integrated cross section at $\sqrt{s}$ =
1960 GeV is: 6.6 $\mu$b
for the exact formula, 2.4 $\mu$b for the simplified formula (see
Eq.~(\ref{KMR-off-diagonal-UGDFs})). For comparison the inclusive
cross section (gluon-gluon component only) is 807 $\mu$b.

In the right panel of Fig.~\ref{figx} we show the differential
cross section in transverse momentum of the charm quark. Compared 
to the inclusive case, the exclusive contribution falls significantly 
faster with transverse momentum than in the inclusive case.

\begin{figure}[h!]
\includegraphics[width=0.45\textwidth]{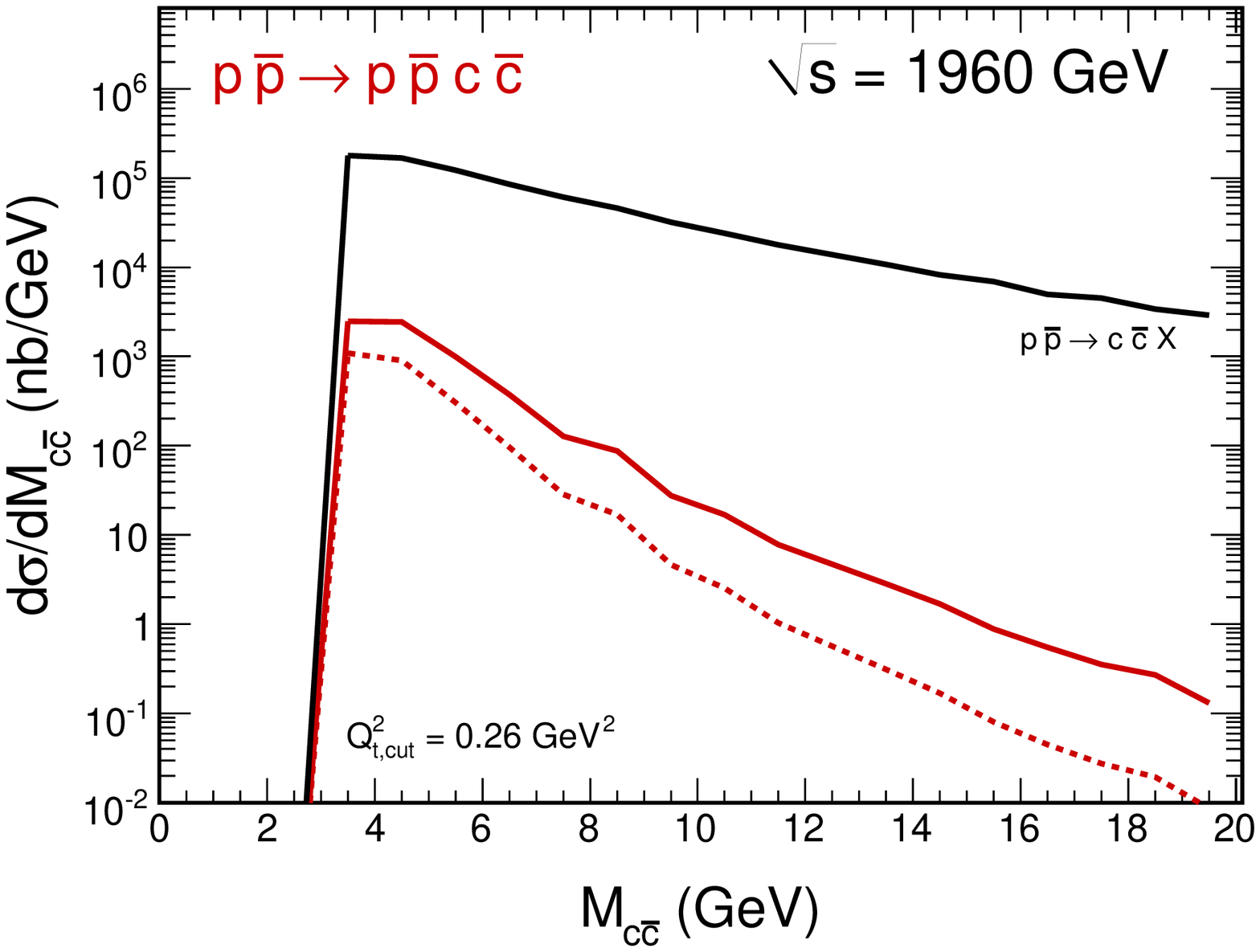}
\includegraphics[width=0.45\textwidth]{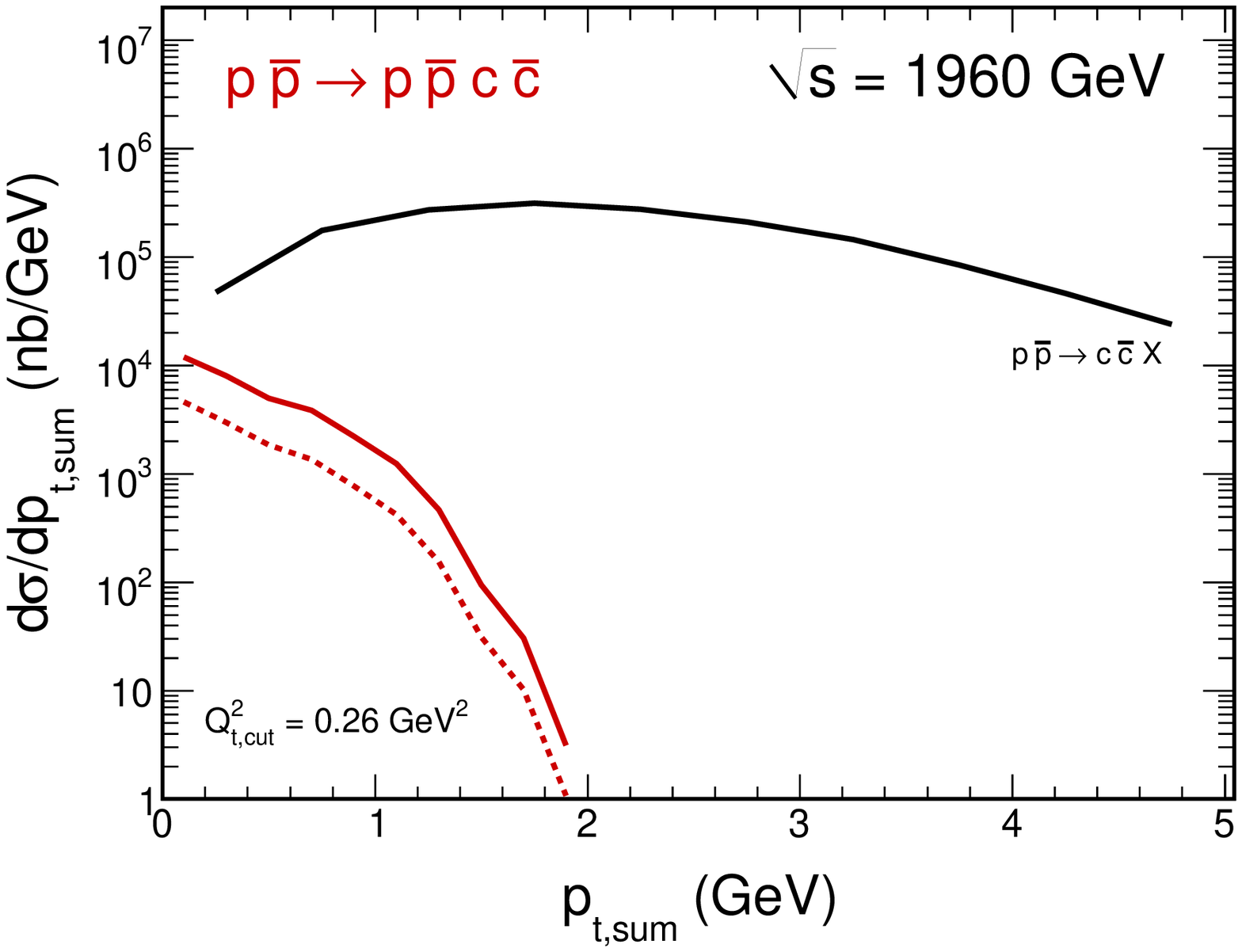}
\caption{\small Invariant mass distribution of the $c \bar c$ pair (left)
and distribution in the transverse momentum of the $c \bar c$ pair (right).
The other details are the same as in the previous figure 
} 
\label{fig:dsig_dm34dptsum}
\end{figure}

In Fig.~\ref{fig:dsig_dm34dptsum} we show the distribution in the
invariant mass of the $c$ $\bar c$ system. 
Compared to the inclusive case the invariant mass distribution for 
the EDD component is significantly steeper. 
This is due to the Sudakov-like form factor
which, according to the procedure described above, damps the cross
section for large invariant masses.

As in the inclusive case within the $k_t$-factorisation approach
the $c \bar c$ pair possesses the transverse momentum
different from zero. The corresponding distribution is shown in
the right panel of Fig.~\ref{fig:dsig_dm34dptsum}. The distribution for 
the exclusive case (the two lower lines) is much narrower compared to 
the inclusive case (the upper line).

\subsection{$pp \to pp b \bar b$ }

In parallel to the exclusive $b \bar b$ production, we calculate the
differential cross sections for exclusive Higgs boson production. 
Compared to the standard KMR approach here we calculate 
the amplitude with the hard subprocess
$g^*g^*\to H$ taking into account off-shellness of the active
gluons.
The details of the off-shell matrix element can be found
in Ref.~\cite{PTS_ggH_vertex}.
In contrast to the exclusive
$\chi_c$ production \cite{PST_chic}, due to a large factorization
scale $\sim M_H$ the off-shell effects for 
$g^*g^*\to H$ give only a few percents to the final result. 

The same unintegrated gluon distributions based on 
the collinear distributions are used for 
the Higgs and continuum $b{\bar b}$ production. 
In the case of exclusive Higgs production we calculate the
four-dimensional distribution in the standard kinematical variables:
$y, t_1, t_2$ and $\phi$.
Assuming the full coverage for outgoing protons
we construct the two-dimensional distributions 
$d \sigma / dy d^2 p_t$ in Higgs rapidity and transverse momentum. 
The distribution is used then in a simple Monte Carlo code 
which includes the Higgs boson decay into the $b {\bar b}$ channel. 
It is checked subsequently whether $b$ and $\bar b$
enter into the pseudorapidity region spanned by the central detector.

\begin{figure}
\begin{center}
\includegraphics[width=8cm]{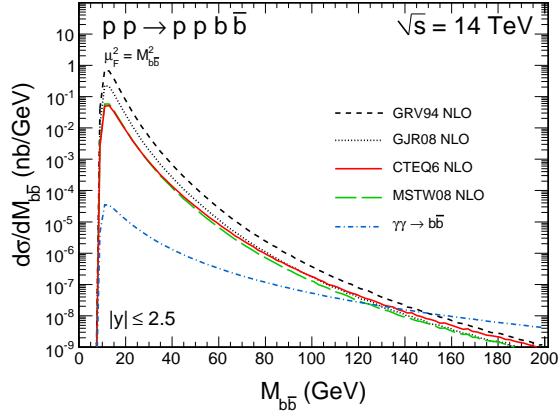}
\end{center}
\caption{ 
The $b \bar b$ invariant mass distribution for $\sqrt{s}$ =
14 TeV and for $-2.5 < y < 2.5$ corresponding 
to the ATLAS/CMS detectors. 
The absorption effects were taken into account by multiplying 
by the gap survival factor $S_G$ = 0.03.}
\label{fig:dsigma_dMbb_PDFs}
\end{figure}

\begin{figure}
\includegraphics[width=7.0cm]{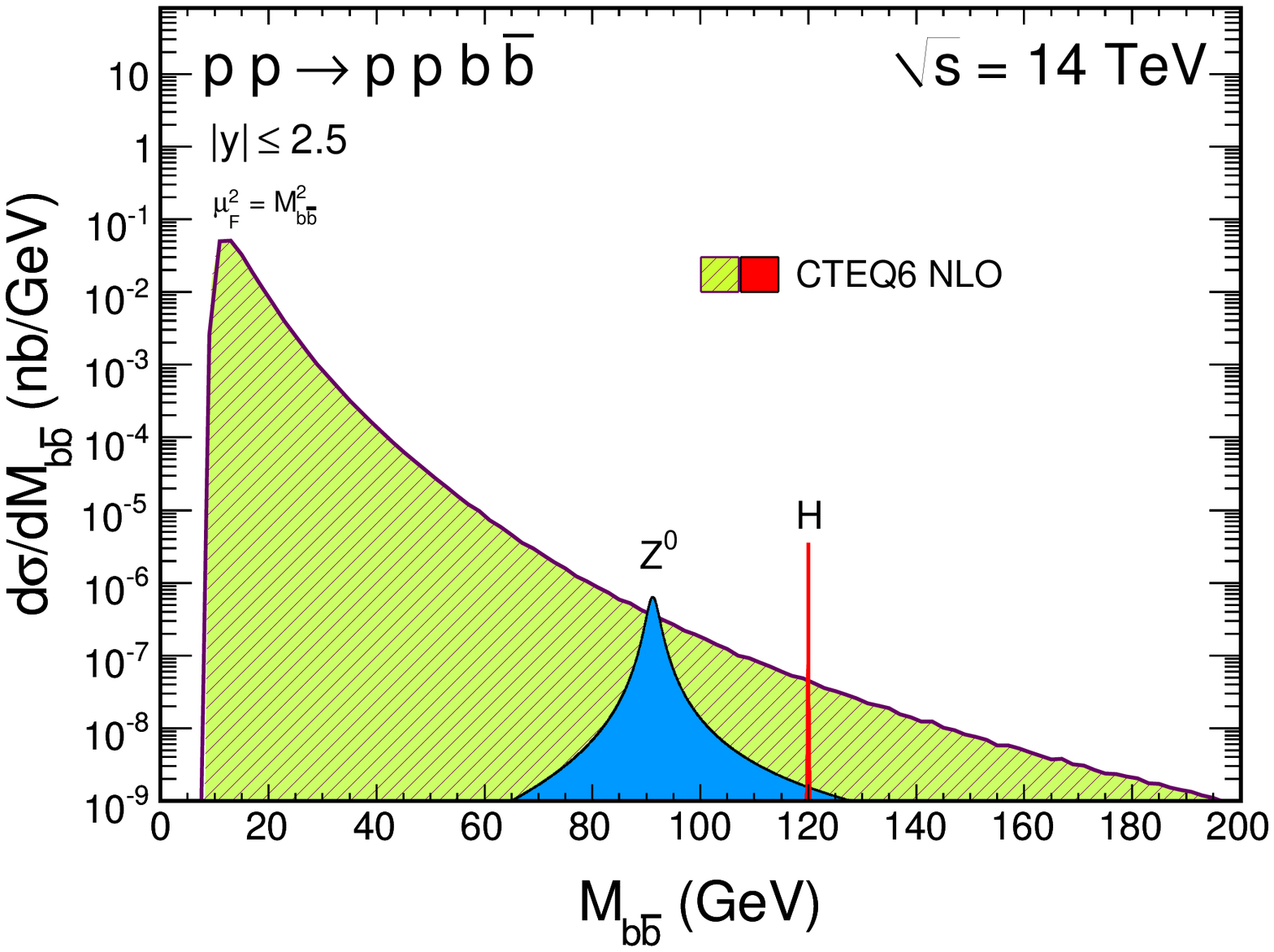}
\includegraphics[width=7.0cm]{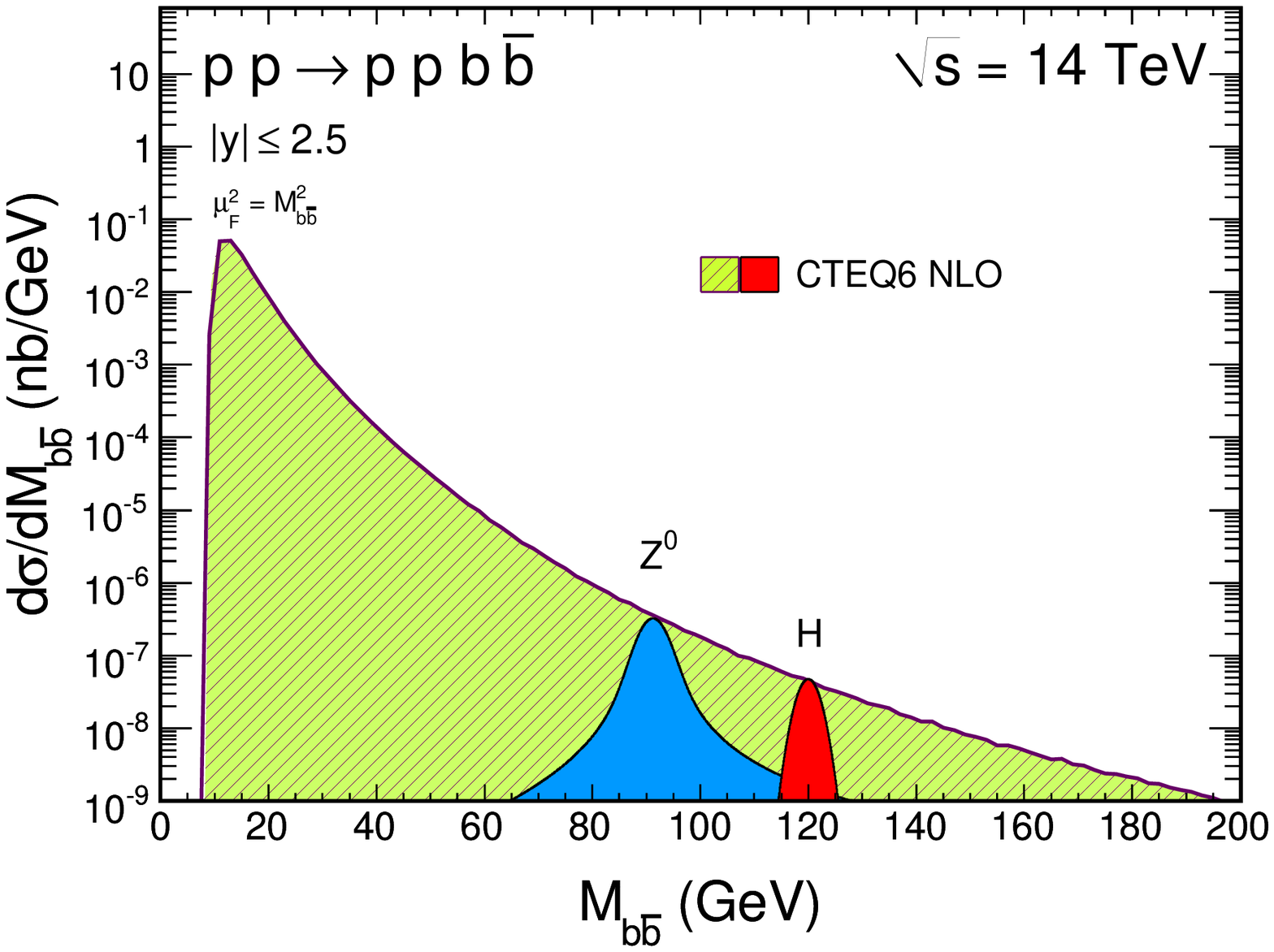}
\caption{The $b \bar b$ invariant mass distribution for $\sqrt{s}$ =
14 TeV and for $b$ and $\bar b$ jets in the rapidity interval $-2.5
< y < 2.5$ corresponding to the ATLAS detector. The absorption
effects for the Higgs boson and the background were taken into
account by multiplying by the gap survival factor $S_G$ = 0.03. The
left panel shows purely theoretical predictions, while the right
panel includes experimental effects due to experimental uncertainty
in invariant mass measurement.} 
\label{fig:dsigma_dMbb_fully}
\end{figure}

In Fig.~\ref{fig:dsigma_dMbb_PDFs} we show the most essential
distribution in the invariant mass of the centrally 
produced $b \bar b$ pair, which is also being the missing mass of 
the two outgoing protons. In this calculation we have taken 
into account typical detector limitations in rapidity 
$-2.5<y_{b},y_{\bar b}<2.5$.
We show results with different collinear gluon distributions
from the literature: GRV \cite{GRV94}, CTEQ \cite{CTEQ}, GJR \cite{GJR}
and MSTW \cite{MSTW}. The results obtained with radiatively generated
gluon distributions (GRV, GJR) allow to use low values of
$Q_t = q_{0t}, q_{1t}, q_{2t}$ whereas for other gluon distributions
an upper cut on $Q_t$ is necessary. 
The integrated double-diffractive $b \bar b$ contribution calculated 
here seems bigger than the contribution of the exclusive photoproduction 
of $b \bar b$ estimated in \cite{GM07}.
The lowest curve in Fig.\ref{fig:dsigma_dMbb_PDFs} represents 
the $\gamma \gamma$ contribution \cite{MPS2010_bbbar}.
While the integrated over phase space $\gamma \gamma$ contribution
is rather small, it is significant compared to
the double-diffractive component at large $M_{b \bar b} >$ 100 GeV.
This can be understood by a damping of the double diffractive
component at large $M_{b \bar b}$ by the Sudakov form factor 
\cite{KMR_Higgs,MPS10}.
In addition, in contrast to the double-diffractive component
the absorption for the $\gamma \gamma$ component is very small
and in practice can be neglected.

In the left panel of Fig.\ref{fig:dsigma_dMbb_fully} we show the 
double diffractive contribution for a selected (CTEQ6 \cite{CTEQ}) 
collinear gluon distribution and the contribution from 
the decay of the Higgs boson 
including natural decay width calculated as in 
Ref.~\cite{Passarino_decay_width}, see
the sharp peak at $M_{b \bar b}$ = 120 GeV.
The phase space integrated cross section for the Higgs production,
including absorption effects with $S_G = 0.03$ 
is somewhat less than 1 fb.
The result shown in Fig.\ref{fig:dsigma_dMbb_fully} includes 
also the branching fraction for BR($H \to b \bar b) \approx$ 0.8 
and the rapidity restrictions. The second much broader 
Breit-Wigner type peak corresponds to the exclusive production 
of the $Z^0$ boson with the cross section 
calculated as in Ref.~\cite{CSS09}. The exclusive cross section
for $\sqrt{s}$ = 14 TeV is 16.61 fb including absorption.
The branching fraction BR($Z^0 \to b \bar b) \approx$ 0.15 
has been included in addition. In contrast to the
Higgs case the absorption effects for the $Z^0$ production are much
smaller \cite{CSS09}. The sharp peak
corresponding to the Higgs boson clearly sticks above the
background. In the above calculations we have assumed an ideal
no-error measurement.

In reality the situation is, however, much worse as both protons 
and in particular $b$ and $\bar b$ jets are measured with a certain
precision which automatically leads to a smearing in $M_{b \bar b}$ .
Experimentally instead of $M_{b \bar b}$ one will measure rather
two-proton missing mass ($M_{pp}$). 
The experimental effects are included in the simplest way by a
convolution  of the theoretical distributions with the Gaussian smearing
function $
G(M) = \frac{1}{\sqrt{2 \pi} \sigma}
\exp\left( \frac{ (M-M_H)^2 } { 2 \sigma^2 } \right)$
with $\sigma$ = 2 GeV \cite{Pilkington_private,Royon_private}
which is determined mainly by the precision of measuring forward
protons.
In the right panel we show the two-proton missinging mass distribution 
when the smearing is included. 
Now the bump corresponding to the
Higgs boson is below the $b \bar b$ background. With the
experimental resolution assumed above the identification of the
Standard Model Higgs seems rather difficult. The situation for
some scenarios beyond the Standard Model may be better
\cite{HKRSTW08,Pilkington}.

\begin{figure}
\begin{center}
\includegraphics[width=5.0cm]{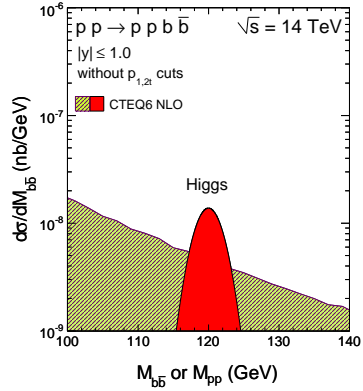}
\end{center}
\caption{The $b \bar b$ invariant mass distribution for $\sqrt{s}$ =
14 TeV for a limited range of $b$ and $\bar b$ rapidities: $-1 <
y < 1$.}
\label{fig:dsigma_dMbb_cuts}
\end{figure}

Can the situation be improved by imposing
further cuts? In Fig.~\ref{fig:dsigma_dMbb_cuts} (left panel) we show the
result for a more limited range of $b$ and $\bar b$
rapidity, i.e. not making use of the whole coverage of the
main LHC detectors. Here we omit the $Z^0$ contribution and
concentrate solely on the Higgs signal. Now the signal-to-background
ratio is somewhat improved. This would be obviously at the expense
of a deteriorated statistics. Similar improvements of the signal-to-background
ratio can be obtained by imposing cuts on jet transverse momenta.
Detailed studies of the role of cuts will be discussed in \cite{MPS_Higgs_large}.

\vspace{0.5cm}

We are indebted to Valery Khoze, Misha Ryskin, Andy Pilkington and
Christophe Royon for a discussion and exchange of useful information.


\end{document}